\documentstyle[12pt,bezier]{article}
\newfont{\mathea}{msam10 scaled\magstep0}
\newfont{\matheb}{msbm10 scaled 1095}
\newfont{\tmpEins}{cmsy10 scaled 2074}
\newfont{\tmpZwei}{cmsy10 scaled 1095}
\newfont{\tmpDrei}{cmsy10 scaled 1000}
\newfont{\tmpVier}{cmsy5 scaled 1000}
\newfont{\tmpFuenf}{msbm7 scaled\magstep0}

\def\Bbb#1{\mathchoice{\mbox{\matheb #1}}{\mbox{\matheb #1}}%
 {\mbox{\tmpFuenf #1}}{\mbox{\tmpFuenf #1}}}

\def\restriction{\mathchoice{
 \mbox{\unitlength1cm\begin{picture}(.2,.4)%
  \bezier{5}(.07,.3)(.1,.27)(.13,.24)%
  \put(.07,.35){\line(0,-1){.5}}\end{picture}}}{
 \mbox{\unitlength1cm\begin{picture}(.2,.4)%
  \bezier{5}(.07,.3)(.1,.27)(.13,.24)%
  \put(.07,.35){\line(0,-1){.5}}\end{picture}}}{
 \mbox{\mathea\symbol{22}}}{
 \mbox{\mathea\symbol{22}}}}

\def\dach#1#2{\mbox{$\mathop{\vbox{\ialign{%
  $##\crcr\hfil #1 \hfil$\crcr}}}\limits^{\scriptscriptstyle #2}$}}

\def\rnzs{\dach{\rho_2}{\mbox{$\scriptscriptstyle\kern-.7mm0$}}\kern-1.2mm'}

\def\Subset{\mbox{$\subset\kern-.5mm\subset$}}
\newcommand{\LI}{\mbox{{\rm L$^{\kern-.15em\raise.2ex\hbox{\scriptsize 1}}$}}}

\def\Ldummy{\left.\bgroup}
\def\Rdummy{\egroup^{\rule{0mm}{1.4mm}}\right.}
\def\LA{\left\langle\bgroup}
\def\RA{\egroup^{\rule{0mm}{1.4mm}}\right\rangle_{\cal A}^{}}
\def\LR{\left(\bgroup}
\def\RR{\egroup^{\rule{0mm}{1.4mm}}\right)}
\def\LG{\left\{\bgroup}
\def\RG{\egroup^{\rule{0mm}{1.4mm}}\right\}}

\def\Wort#1{\mbox{{\rm #1\kern.1em}}}

\def\lfac#1#2{\vcenter{\hbox{$#1\kern-.2em\raise-.6ex\hbox{\Large{/}}%
 \kern-.2em\raise-1.2ex\hbox{$#2$}$}}}

\def\gin{\mbox{\tmpZwei\symbol{91}\kern-1.4mm\rule{.2mm}{1.85mm}\kern1.4mm}}
\def\gni{\mbox{\tmpZwei\symbol{92}\kern-1.4mm\rule[.15mm]{.2mm}{1.85mm}%
  \kern1.4mm}}

\def\EINS{{\mathop{1\kern-.25em\mbox{{\rm{\small l}}}}}}

\begin{document}

\Large Decay Semigroups for the Resonances of
 Quantum \\Mechanical Scattering 
Systems  

\vspace{3mm}

\normalsize Hellmut Baumg\"artel

\vspace{3mm}

Mathematical Institute, University of Potsdam

Am Neuen Palais 10, PF 601553

D-14415 Potsdam, Germany

e-mail: baumg@uni-potsdam.de

\vspace{3mm}

\begin{abstract}
For selected classes of quantum mechanical Hamiltonians a canonical
association of a decay semigroup is presented. The spectrum of the generator of this
semigroup is a pure eigenvalue spectrum and it coincides with the set of all
resonances. The essential condition for the results is the meromorphic continuability
of the scattering matrix onto $\Bbb{C}\setminus(-\infty,0]$ and the rims
$\Bbb{R}_{-}\pm i0$. Further finite multiplicity is assumed. 
The approach is based on an adaption of the Lax-Phillips scattering theory to
semi-bounded Hamiltonians. It is applied to trace class perturbations with
analyticity conditions. A further example is
the potential scattering for central-symmetric potentials with compact support 
and angular momentum $0$. 
\end{abstract}

\vspace{3mm}


\vspace{3mm}


\vspace{3mm}

\section{Introduction}

The basic topic of this paper is the mathematical theory of quantum mechanical resonances
which can be traced back to the origin of scattering theory in quantum mechanics.
In quantum scattering systems bumps in cross sections often can be described
by Breit-Wigner formulas like
$E\rightarrow c((E-E_{0})^{2}+(\Gamma/2)^{2})^{-1}$
which associate a "resonance" at energy $E$ with halfwidth $\Gamma/2$. If the scattering
matrix $E\rightarrow S(E)$ is assumed to have a 
simple pole $E_{0}-i\Gamma/2$ in the lower half plane near the real axis then the scattering 
cross
section can be approximately described by a Breit-Wigner formula . Therefore resonances are
associated with poles of the analytic continuation of the scattering matrix into the
lower half plane across the positive half line and usually these poles are called resonances
(see e.g. [1,2,3]).
In principle, the investigation of the resonances requires
explicit knowledge of the scattering matrix $S(\cdot)$. It depends on the Hamiltonian $H$,
if the so-called "unperturbed" Hamiltonian is assumed to be fixed. In general it  
is difficult to obtain properties of the analytic structure of $S(\cdot)$ directly.
A second difficulty is to assign to such a pole in a rigorous
way a "state of finite lifetime" such that it is an eigenstate to that pole as eigenvalue
of a (non-selfadjoint) operator closely related to the quantum Hamiltonian. These facts
caused a distinguished history of the theory of resonances.

\vspace{3mm}

A well-known approach, 
the so-called Aguilar-Balslev-Combes-Simon(ABCS)-theory (see e.g. [4,5,6]) 
starts with a modification of the definition of resonances to 
be the poles of the analytic continuation of all matrix elements 
$(f,R_{H}(z)g),\,R_{H}(z)$ the resolvent of $H,\,f,g\in{\cal A}$ a dense set of vectors in 
the Hilbert space, from the upper half plane across the positive half line into the lower
half plane. This modification is due to the fact that there is a connection
between the scattering matrix and the resolvent of the Hamiltonian $H$ such that in 
many cases
the poles of these matrix elements can be shown to be identical with the resonances.
The approach aims then essentially at Schr\"odinger Hamiltonians $H$, for example on
$L^{2}(\Bbb{R}^{3})$, and their dilatation
transforms $U(\Theta)HU(\Theta)^{-1}$, which can be controlled explicitly, where
the dilatation transformation is given by
$(U(\Theta)f)(r):=e^{3\Theta/2}f(e^{\Theta}r)$.
Then the associated non-selfadjoint operator is given by the 
dilatation transform  of $H$. 

\vspace{3mm}

A second approach aims to the characterization of the resonances by spectral 
properties of $H$ directly. It is successful if there is a special
holomorphic operator function $F(\cdot)$, defined on $\Bbb{C}\setminus [0,\infty)$, 
depending on $H$ such that the negative (real) eigenvalues $\lambda$ of $H$
are characterized by the condition $\mbox{ker}\,F(\lambda)\supset\{0\}$. For example
this is true for trace class perturbations. If $F(\cdot)$ can be analytically continued into
the lower half plane across the positive real line then possible positive eigenvalues and
also the resonances $\zeta$ are characterized by the same condition
$\mbox{ker}\,F(\zeta)\supset\{0\}$. This shows the close relationship of resonances
to eigenvalues (see e.g. [7], see also [8]). It can be improved by construction
of an extension of $H$ by means of an appropriate Gelfand triplet. Then the resonances
appear as the eigenvalue spectrum of this extension, where the mentioned (continued)
condition leads to a kind of "boundary condition" for the generalized eigenvalue 
problem. As a by-product this approach solves the problem to associate the right 
Gamov vectors to resonances (for the concept {\em Gamov vector} see [9]), 
because the corresponding
eigenvectors of a resonance $\zeta$ turn out to be specific ones whose coefficients
are well-defined by the (continued) characterization condition for $\zeta$ (see [10],
[11]). 

\vspace{3mm}

A third approach is based on the application of the famous Lax-Phillips(LP)-scattering theory 
[12] to quantum mechanical resonance problems. At
first sight it seems that this theory is unsuitable for such an application  because the 
Hamiltonians in this theory have necessarily absolutely continuous
spectrum of constant multiplicity coinciding with the whole real line. However, the advantage
in the "classical" LP-theory (where outgoing and incoming subspaces are orthogonal)
is that the second difficulty, mentioned above, is overcome in a most
convincing way, because in this case the associated non-selfadjoint operator $A$
is the generator of a strongly continuous contractive semigroup for $t\geq 0$ tending to
$0$ for $t\rightarrow\infty$ such that the set of all resonances coincides with
the eigenvalue spectrum of $A$ in the lower half plane and the corresponding eigenvectors
for a resonance $\zeta$
are given explicitly by vectors of the form $E\rightarrow k(E-\zeta)^{-1}$ (Gamov vectors),
where $k$ is
from the multiplicity Hilbert space and depends on the scattering matrix.

\vspace{3mm}

The general LP-theory can be considered as a special part of the abstract scattering
theory, where the unperturbed operator is the multiplication operator $M$ on the
Hilbert space $L^{2}(\Bbb{R},{\cal K},d\lambda)$ on the whole real line and ${\cal K}$
is the (separable) multiplicity Hilbert space, $\dim{\cal K}\leq\infty$; the
admissible Hamiltonians generate unitary evolutions with the well-known outgoing and incoming
subspaces. On the other hand, the Hamiltonians $H$ of quantum systems are often
semibounded, where the absolutely continuous spectrum is $[0,\infty)$ with 
constant multiplicity such that the unperturbed Hamiltonian, without loss of generality, can
be assumed to be the multiplication operator $M_{+}$ on a Hilbert space 
$L^{2}(\Bbb{R}_{+},{\cal K},d\lambda)$. These facts suggest the question under which
conditions on quantum mechanical scattering systems $\{H,M_{+}\}$ the characterization
problem for resonances can be solved by a modification or adaption of the successful
methods of the LP-theory.

\vspace{3mm}

There are also other approaches for the application of the "classical" LP-theory to quantum
resonance problems. An example is the approach of Strauss-Horwitz-Eisenberg ([13], see also
[14]). They consider the "physical" Hilbert space ${\cal H}$ of quantum mechanics as the
multiplicity Hilbert spaces ${\cal K}_{t}$ of the LP-theory, where ${\cal K}_{t}:=
U(t){\cal H}$ and $U(\cdot)$ is the unitary quantum evolution, where it is used that
the multiplicity Hilbert spaces ${\cal K}_{t}$ are all isomorphic.

\vspace{3mm}

This paper presents a solution
of the problem to associate  
to the given Hamiltonian $H$ and to its resonances 
a non-selfadjoint operator such that the eigenvalue spectrum
of this operator coincides with the set of all resonances by an adaption of methods
of the general LP-theory (see [12]).
In fact, like in the "classical" LP-theory, this operator
is the generator of a so-called {\em decay semigroup} associated to $H$.
The focus of the consideration is a pure conceptual one, there are no direct 
computational consequences.
The procedure is carried out for Hamiltonians $H$ satisfying the following basic properties:
\begin{itemize}
\item [(i)]
$H$ is semi-bounded with absolutely continuous spectrum coinciding with
$[0,\infty)$ and of constant finite multiplicity,
\item [(ii)]
together with a so-called "free" Hamiltonian it forms an asymptotically complete scattering
system whose scattering matrix
is
meromorphically continuable into $\Bbb{C}\setminus (-\infty,0]$ and it is meromorphic
also on the rims $\Bbb{R}_{-}\pm i0$. 
\end{itemize}

\vspace{3mm}

First, in Section 2, it is pointed out that it is sufficient to solve the problem for
Hamiltonians $H$ whose absolutely continuous part together with the "free" Hamiltonian
$M_{+}$, the multiplication operator on a Hilbert space 
${\cal H}_{+}:=L^{2}(\Bbb{R}_{+},{\cal K})$,
form an asymptotically complete scattering system, 
where ${\cal K}$ is the multiplicity Hilbert space. It is assumed
that $\dim{\cal K}<\infty$. A theorem of Wollenberg is quoted which ensures that there are
no existence problems w.r.t. condition (ii).

Second, in Section 3, several derived spectral invariants of
$M_{+}$ are introduced, for example the so-called time-asymptotic(TA-) semigroups and their
adjoints, one of them is called the {\em characteristic semigroup}.
This step uses a distinguished isometry between ${\cal H}_{+}$ and the corresponding
Hardy space ${\cal H}^{2}_{+}(\Bbb{R},{\cal K})$.
The first result in this context is due to [15]. The construction of the isometry
uses generalizations and improvements due to [16] and [17] and the polar decomposition
of bounded operators.

The decisive step, the basic idea of the adaption, is described in Section 4:
The association of an invariant subspace of the characteristic semigroup,
depending on the scattering matrix of the scattering system $\{H,M_{+}\}$.
In this connection the "classical" LP-case appears only as a special case, the constructions
require more general invariant subspaces of the characteristic semigroup.
Its restriction to this invariant subspace is a strongly continuous contractive semigroup
which tends strongly to zero for $t\rightarrow\infty$, whose generator $B_{+}$ depends only on
the scattering matrix $S(\cdot)$.

In Section 5 the spectrum of $B_{+}$ is calculated under slightly different additional
assumptions. In every case the result is that 
the spectrum
of $B_{+}$ is a pure eigenvalue spectrum which coincides with the set of all
resonances. 
The cases $S(\lambda+i0)=S(\lambda-i0)$ and $S(\lambda+i0)\neq S(\lambda-i0)$
for $\lambda<0$ are treated separately because they are completely different. 
As an
example for the second case the potential scattering for a central-symmetric potential
with compact support and angular momentum $l=0$ is considered (for this example see
[10]).

In Section 6 it is shown that a class of trace-class perturbations with 
special analyticity properties satisfy the assumptions of Theorem 3. In this case
the condition $\dim{\cal K}<\infty$ is dipensable. Also special generalized
Friedrichs models, considered in [11], satisfy these assumptions.

\section{Asymptotically complete scattering systems}

In the following $H$ and $H_{0}$ are selfadjoint operators (Hamiltonian and unperturbed
Hamiltonian) on a separable Hilbert space
${\cal H}$, which are semibounded with absolutely continuous spectrum which coincides 
with $[0,\infty)$ and of constant multiplicity $m,\,1\leq m\leq\infty$. The absolutely
continuous subspaces are denoted by $P^{ac}{\cal H}$ and ${\cal H}_{0}$, respectively.
Then $H_{0}\restriction{\cal H}_{0}$ is unitarily equivalent to the multiplication 
operator $M_{+}$
on the Hilbert space ${\cal H}_{+}:= L^{2}(\Bbb{R}_{+},{\cal K})$, where $\dim{\cal K}=m$.
 Let $\Phi$ on ${\cal H}$ and $\Phi_{0}$ on ${\cal H}_{0}$ be unitaries realizing
spectral representations of $H$ and $H_{0}\restriction{\cal H}_{0}$, respectively.
Then$\Phi_{0} e^{itM_{+}}=e^{itH_{0}}\Phi_{0},\,-\infty<t<\infty$ holds.
If $\{H,H_{0}\}$ is an asymptotically complete scattering system then the wave operators
$\mbox{s-lim}_{t\rightarrow \pm\infty}e^{itH}e^{-itH_{0}}$ exist and are isometries from
${\cal H}_{0}$ onto $P^{ac}{\cal H}$. The (unitary) scattering
operator $S$ acts on ${\cal H}_{0}$. Then $\Phi_{0}S\Phi_{0}^{-1}$ acts on ${\cal H}_{+}$
and commutes with $e^{itM_{+}}$, i.e. it acts by the (unitary)
scattering matrix $\Bbb{R}_{+}\ni\lambda\rightarrow S(\lambda)\in{\cal L}({\cal K})$ 
on ${\cal H}_{+}$ and the operators $S(\lambda)$ are unitary.
$\Phi\Phi_{0}^{-1}$ is an isometry from 
${\cal H}_{0}$ onto $P^{ac}{\cal H}$. Then the systems 
$\{\Phi^{-1}H\Phi,\Phi_{0}^{-1}H_{0}\Phi_{0}\}$ and
$\{\Phi^{-1}H\Phi,M_{+}\}$ coincide because of $M_{+}=\Phi_{0}^{-1}H_{0}\Phi_{0}$.
The scattering operator of this system coincides with the S-matrix-function. 
Therefore, without loss
of generality we can restrict the consideration to scattering systems $\{H,M_{+}\}$
acting on ${\cal H}_{+}\oplus{\cal E}$,
where ${\cal E}$ corresponds to the space ${\cal H}\ominus{\cal H}_{0}$. 
Recall for these systems the solution of the inverse problem:

\vspace{3mm}

THEOREM 1 (Wollenberg). {\em To every unitary operator} $S$ {\em on} ${\cal H}_{+}$ {\em with}
\[
Se^{itM_{+}}=e^{itM{+}}S,\quad -\infty<t<\infty,
\]
{\em there is a selfadjoint operator} $H$ {\em on} ${\cal H}_{+}$ {\em such that}
$\{H,M_{+}\}$ {\em is an asymptotically complete scattering system whose scattering
operator coincides with} $S$.

\vspace{3mm}

For the proof see [18] (see also [19]). The description of all solutions of the inverse
problem (see e.g. [19]) shows that $M_{+}$ and $S$ form a complete system of spectral
invariants of the Hamiltonian $H$ in this context. Theorem 1 ensures that
to every unitary operator $S$ whose scattering matrix satisfies condition (ii)
of Section 1 there is a corresponding selfadjoint operator $H$.

\section{Derived spectral invariants of $M_{+}$}

\subsection{Semigroups on the Hardy spaces}
First we collect some basic facts on Hardy spaces and fix notation.
Let ${\cal H}:= L^{2}(\Bbb{R},{\cal K},d\lambda)$ and ${\cal H}^{2}_{\pm}:=
{\cal H}^{2}_{\pm}(\Bbb{R},{\cal K})$ the Hardy subspaces.
The projections $Q_{\pm}$ onto ${\cal H}^{2}_{\pm}$ are given by
\[
{\cal H}\ni f\rightarrow Q_{\pm}f(z):=\pm\frac{1}{2\pi i}\int_{-\infty}^{\infty}
\frac{f(\lambda)}{\lambda-z}d\lambda,\quad z\in\Bbb{C}_{\pm}.
\]
Further let $P_{\pm}$ be the projections
\[
{\cal H}\ni f\rightarrow P_{\pm}f(\lambda)=\chi_{\Bbb{R}_{\pm}}(\lambda)f(\lambda).
\]
Then $P_{\pm}{\cal H}=L^{2}(\Bbb{R}_{\pm},{\cal K},d\lambda)=:{\cal H}_{\pm}.$
We use the Fourier transformation in the form
\[
Ff(\lambda):=\frac{1}{\sqrt{2\pi}}\int_{-\infty}^{\infty}e^{-i\lambda x}f(x)dx.
\]
Then
\[
Q_{\pm}=FP_{\mp}F^{-1}.
\]
Next we introduce the so-called shift evolution on ${\cal H}:$
\[
(T(t)g)(x):=g(x-t),\quad g\in{\cal H}.
\]
The subspaces ${\cal H}_{\mp}=P_{\mp}{\cal H}$ are in/out subspaces for $T$:
\[
T(t){\cal H}_{-}\subseteq{\cal H}_{-},\quad t\leq 0,
\]
and
\[
T(t){\cal H}_{+}\subseteq {\cal H}_{+},\quad t\geq 0.
\]
The spectral representation $\hat{T}$ of $T$ is realized by the Fourier transformation:
\[
\hat{T}(t):=FT(t)F^{-1},\quad t\in\Bbb{R}.
\]
We use the denotation $\hat{T}(t)=e^{-itM}$ where $M$ is the multiplication
operator on ${\cal H}$,
\[
Mf(\lambda)
:=\lambda f(\lambda),\quad f\in{\cal H}.
\]
$M$ is the generator of the (unitary) group $\hat{T}(\cdot)$. The same denotation 
we use in the following
for the contractive semigroups which occur in the paper to indicate its generator.
Therefore, the subspaces $Q_{\pm}{\cal H}={\cal H}^{2}_{\pm}$ are in/out subspaces for
$e^{-itM}$:
\[
e^{-itM}{\cal H}^{2}_{+}\subseteq{\cal H}^{2}_{+},\quad t\leq 0,
\]
and
\[
e^{-itM}{\cal H}^{2}_{-}\subseteq{\cal H}^{2}_{-}\quad t\geq 0.
\]
In other words, we have the relations
\begin{equation}
e^{-itM}Q_{+}=Q_{+}e^{-itM}Q_{+}\quad t\leq 0,
\end{equation}
and
\[
e^{-itM}Q_{-}=Q_{-}e^{-itM}Q_{-}\quad t\geq 0.
\]
This means that the unitary evolution group
\[
\Bbb{R}\ni t\rightarrow e^{-itM}
\]
generates in the Hardy subspaces ${\cal H}^{2}_{\pm}$ strongly continuous isometric
semigroups
\begin{equation}
e^{-itM}\restriction {\cal H}^{2}_{+}=e^{-itA_{+}},\quad t\leq 0,
\end{equation}
and
\begin{equation}
e^{-itM}\restriction {\cal H}^{2}_{-}=e^{-itA_{-}},\quad t\geq 0,
\end{equation}
where the operators $A_{\pm}$, defined on the Hardy 
spaces ${\cal H}^{2}_{\pm}$, are the generators of these semigroups. Their spectral
structure is well-known:

\vspace{2mm}

PROPOSITION 1. {\em The generator} $A_{\pm}$ {\em is maximally symmmetric, i.e. there 
is no symmetric extension of} $A_{\pm}$. {\em It satisfies the following
properties}:
\begin{itemize}
\item[(i)]
$\mbox{dom}\,A_{\pm}=\{f\in\mbox{dom}\,M\cap{\cal H}^{2}_{\pm}: Mf\in{\cal H}^{2}_{\pm}\}$
{\em and}
\[
(A_{\pm}f)(z)=zf(z),\quad z\in\Bbb{C}_{\pm},\quad f\in\mbox{dom}\,A_{\pm}.
\]
\item[(ii)]
{\em For} $\zeta\in\Bbb{C}_{\pm}$ {\em the image} $(\zeta-A_{\pm})\mbox{dom}\,A_{\pm}$
{\em is a subspace and coincides with}
\[
{\cal N}_{\zeta}:=\{f\in{\cal H}^{2}_{\pm}: f(\zeta)=0\},
\]
\item[(iii)]
{\em The deficiency space}
\[
{\cal D}_{\zeta}:={\cal H}^{2}_{\pm}\ominus {\cal N}_{\zeta}
\]
{\em is given by}
\[
{\cal D}_{\zeta}=\{f\in{\cal H}^{2}_{\pm}: f(z):=\frac{k}{(z-\overline{\zeta})},\quad
k\in{\cal K}\}.
\]
\item[(iv)]
\[
\mbox{spec}\,A_{\pm}=\mbox{clo}\,\Bbb{C}_{\pm},\quad \mbox{res}\,A_{\pm}=\Bbb{C}_{\mp}.
\]
\end{itemize}
For the proof see e.g. [20]. 
Further we need an improvement of relation (ii) of Proposition 1.

\vspace{3mm}

LEMMA 1. {\em Let} $\{(\xi_{j},g_{j}), j=1,...r\}\subset\Bbb{C}_{+}\times\Bbb{N}$
{\em and} ${\cal N}_{\xi,g}:=
\{f\in{\cal H}^{2}_{+}: f^{(m_{j})}(\xi_{j})=0,\,m_{j}=1,2,...g_{j},
\,j=1,2,...r\}.$ {\em Then}
\begin{itemize}
\item[(i)]
${\cal N}_{\xi,g}\subset{\cal H}^{2}_{+}$ {\em is a subspace and}
\item[(ii)]
{\em the orthogonal complement} ${\cal H}^{2}_{+}\ominus {\cal N}_{\xi,g}$ {\em of} 
${\cal N}_{\xi,g}$ 
{\em is given by}
\[
\mbox{clo spa}\,\{f\in{\cal H}^{2}_{+}:
f(z):=\frac{k}{(z-\overline{\xi_{j}})^{m_{j}}},\,m_{j}=1,2,...g_{j},\,j=1,2...,r,\,
k\in{\cal K}\}.
\]
\end{itemize}
 
\vspace{3mm}

The proof follows closely that of [16] using the more general identity
\[
\int_{-\infty}^{\infty}(\frac{k}{(x-\overline{\xi})^{m}},g(x))dx=
\int_{-\infty}^{\infty}\frac{1}{(x-\xi)^{m}}(k,g(x))dx=
\frac{2\pi i}{m!}(k,g^{(m-1)}(\xi))
\]
for $g\in{\cal H}^{2}_{+}$.

\vspace{3mm}

Later we use only the semigroup (2) in the form
\begin{equation}
t\rightarrow T_{+}(t):=e^{itM}\restriction{\cal H}^{2}_{+}=e^{itA_{+}},\quad t\geq 0.
\end{equation}
Its adjoint semigroup is of special interest. (1) implies
\begin{equation}
Q_{+}e^{-itM}Q_{+}=Q_{+}e^{-itM},\quad t\geq 0.
\end{equation}
Therefore,
\begin{equation}
T_{+}(t)^{\ast}= Q_{+}e^{-itM}\restriction{\cal H}^{2}_{+}=:C_{+}(t).
\end{equation}

\vspace{3mm}

PROPOSITION 2. {\em The semigroup} $\Bbb{R}_{+}\rightarrow C_{+}(t)$ {\em has
the following properties}:
\begin{itemize}
\item[(i)]
{\em It is strongly continuous and contractive,} $C_{+}(t)=e^{-itC_{+}},\;t\geq 0$,
{\em the generator} $C_{+}$ {\em is closed on} ${\cal H}^{2}_{+},\;\mbox{dom}\,C_{+}$
{\em is dense and} $\Bbb{C}_{+}\subset\mbox{res}\,C_{+}$.
\item[(ii)]
$\mbox{dom}\,C_{+}$ {\em consists of all} $f\in{\cal H}^{2}_{+}$ {\em such that there is
an associated} $k\in{\cal K}$ {\em and the function} $g(\lambda):=\lambda f(\lambda)+k$
{\em is from} ${\cal H}^{2}_{+}$. {\em In this case} $(C_{+}f)(\lambda)=
\lambda f(\lambda)+k$.
\item[(iii)]
$C_{+}=A_{+}^{\ast}$,
\item[(iv)]
$C_{+}(t)(f)(z)=\frac{1}{2\pi i}\int_{-\infty}^{\infty}\frac{e^{-it\lambda}}{\lambda-z}
f(\lambda)d\lambda,\quad f\in{\cal H}^{2}_{+}$.
\item[(v)]
{\em One has} $\mbox{s-lim}_{t\rightarrow\infty}e^{-itC_{+}}=0$.
\end{itemize}
For the proof see [20]. We call the semigroup $C_{+}(\cdot)$ the 
{\em characteristic semigroup}. The spectral structure of its generator $C_{+}$
is given in

\vspace{3mm}

PROPOSITION 3. {\em The generator} $C_{+}$ {\em has the following properties}:
\begin{itemize}
\item[(i)]
$\mbox{res}\,C_{+}=\Bbb{C}_{+},$
\item[(ii)]
{\em The eigenvalue spectrum of} $C_{+}$ {\em coincides with} $\Bbb{C}_{-}$, {\em i.e.
a real-valued point cannot be an eigenvalue},
\item[(iii)]
{\em The eigenspace of the eigenvalue} $\zeta\in\Bbb{C}_{-}$ {\em is given by the
subspace}
\[
{\cal E}_{\zeta}:=\left\{f\in{\cal H}^{2}_{+}: f(z):=\frac{k}{z-\zeta},\;k\in{\cal K}\right\},
\]
{\em and one has}
\[
C_{+}(t)f=e^{-it\zeta}f,\quad f\in{\cal E}_{\zeta}.
\]
\end{itemize}
For the proof see e.g. [20].

\subsection{Transfer of the semigroups to ${\cal H}_{+}$}

By means of the projections $P_{+}$ and $Q_{+}$ a distinguished
isometry between ${\cal H}_{+}$ and ${\cal H}^{2}_{+}$ can be constructed.
According to results of Kato [16] and Halmos [17]
the properties 

\begin{itemize}
\item [(i)]
the subspaces $P_{+}{\cal H}$ and $Q_{\pm}{\cal H}$ are subspaces in generic position (in
the sense of Halmos), i.e.
\[
P_{+}{\cal H}\cap Q_{+}{\cal H}=P_{+}{\cal H}\cap Q_{-}{\cal H}=P_{-}{\cal H}\cap
Q_{+}{\cal H}=P_{-}{\cal H}\cap Q_{-}{\cal H}=\{0\},
\]
\item[(ii)]
$\Vert P_{+}-Q_{\pm}\Vert =1$
\end{itemize}
imply that the submanifolds ${\cal M}_{\pm}:=P_{+}{\cal H}^{2}_{\pm}\subset{\cal H}_{+}$
are dense in ${\cal H}_{+}$. Furthermore, they are  
in/out manifolds for $e^{-itM_{+}}$:
\[
e^{-itM_{+}}{\cal M}_{+}\subseteq{\cal M}_{+},\quad t\leq 0
\]
and 
\[
e^{-itM_{+}}{\cal M}_{-}\subseteq{\cal M}_{-}\quad t\geq 0.
\]
Note that the intersection
\[
{\cal M}_{+}\cap{\cal M}_{-}
\]
is infinite-dimensional (whether it is dense in ${\cal H}_{+}$ is an open question). 

\vspace{3mm}

PROPOSITION 4. {\em There is a distinguished isometry between} ${\cal H}^{2}_{+}$ {\em and}
${\cal H}_{+}$ {\em given by the operators}
\[
R: {\cal H}^{2}_{+}\rightarrow{\cal H}_{+},\quad R^{\ast}: {\cal H}_{+}\rightarrow
{\cal H}^{2}_{+},
\]
{\em where}
\begin{equation}
R:=\mbox{sgn}\,(P_{+}Q_{+})\restriction{\cal H}^{2}_{+}.
\end{equation}
{\em and} $\,\mbox{sgn}\,(P_{+}Q_{+})$ {\em means the partial isometry in
the polar decomposition of} $P_{+}Q_{+}$.

\vspace{3mm}

Proof. The polar decomposition of $P_{+}Q_{+}$ in ${\cal H}$ reads
\[
P_{+}Q_{+}=(P_{+}Q_{+}P_{+})^{1/2}\,\mbox{sgn}\,(P_{+}Q_{+}),
\]
and $\mbox{sgn}\,(P_{+}Q_{+})$ is a partial isometry with initial projection $Q_{+}$
and final projection $P_{+}$ (note that $P_{+}{\cal H}^{2}_{+}\subset{\cal H}_{+}$
is dense in ${\cal H}_{+}).\quad \Box$

\vspace{3mm}

Using the isometry $R:{\cal H}^{2}_{+}\rightarrow{\cal H}_{+}$ we transfer the
semigroups
\[
T_{+}(t)=e^{itM}\restriction{\cal H}^{2}_{+},\quad C_{+}(t)=Q_{+}e^{-itM}\restriction
{\cal H}^{2}_{+},\quad t\geq 0
\]
into subgroups on ${\cal H}_{+}$ by the transformation
\begin{equation}
\tilde{T}_{+}(t):=RT_{+}(t)R^{\ast},\quad \tilde{C}_{+}(t):=RC_{+}(t)R^{\ast},\quad t\geq 0,
\end{equation}
such that $\tilde{T}_{+}(\cdot)$ and $T_{+}(\cdot)$, also $\tilde{C}_{+}(\cdot)$ and
$C_{+}(\cdot)$ are unitarily equivalent, as well as the corresponding generators. Further
$\tilde{C}_{+}(\cdot)$ remains the adjoint semigroup of $\tilde{T}_{+}(\cdot)$.

\vspace{3mm}

The semigroups (8) are derived spectral invariants of $M_{+}$. So far the 
scattering matrix $S(\cdot)$ is not yet into the play. In the next section it is shown
that there are invariant subspaces of $\tilde{C}_{+}(\cdot)$ resp. of
$C_{+}(\cdot)$, depending only on $S(\cdot)$ such that the spectrum of the generator
of this restricted semigroup can be characterized by the resonances. Because of
the mentioned unitary equivalence of $\tilde{C}_{+}(\cdot)$ and $C_{+}(\cdot)$ one
can study the spectral invariant properties of these semigroups by the study of
$C_{+}(\cdot)$ which acts on the Hardy space ${\cal H}^{2}_{+}$.

\section{Invariant subspaces of $T_{+}(\cdot)$ and $C_{+}(\cdot)$}

According to Section 2(ii) the scattering matrix $S(\cdot)$ of the scattering system
$\{H,M_{+}\}$ is assumed to be a holomorphic unitary operator function
$\Bbb{R}_{+}\ni\lambda\rightarrow S(\lambda)$ which is meromorphically continuable
into $\Bbb{C}\setminus (-\infty,0]$ including the rims $\Bbb{R}_{-}\pm i0$.

\vspace{3mm}

By ${\cal M}_{+}\subseteq{\cal H}^{2}_{+}$ we denote the linear manifold of all
$f\in{\cal H}^{2}_{+}$ such that there is a function $g\in{\cal H}^{2}_{+}$ with
\begin{equation}
f(z):=S(z)g(z),\quad z\in\Bbb{C}_{+}.
\end{equation}
The linear manifold of all $g\in{\cal H}^{2}_{+}$ satisfying
\begin{equation}
\sup_{y>0}\int_{-\infty}^{\infty}\Vert S(x+iy)g(x+iy)\Vert^{2}_{\cal K}dx<\infty,
\end{equation}
is denoted by ${\cal N}_{+}$.
According to the Paley-Wiener theorem, this means that for the the function (9)
$\mbox{s-lim}_{y\rightarrow +0}\,f(\lambda+iy)=:f(\lambda+i0)$ exists a.e. and it is
from ${\cal H}^{2}_{+}$. That is, one can write briefly, without ambiguity,
${\cal M}_{+}=S{\cal N}_{+}$. For example, if $S(\cdot)$ is
holomorphic on $\Bbb{C}_{+}$ and $\Vert S(z)\Vert\leq 1$ there, then 
${\cal N}_{+}={\cal H}^{2}_{+}$
and ${\cal M}_{+}=S{\cal H}^{2}_{+}$.

\vspace{3mm}

Note that ${\cal N}_{+}$ is invariant w.r.t. the multiplication with $e^{itz}$, i.e.
if $g\in{\cal N}_{+}$, i.e. (10) is true then $z\rightarrow e^{itz}g(z)$ satisfies
(10), too, because
\[
\Vert S(z)\{e^{itz}g(z)\}\Vert_{\cal K}=\Vert e^{itz}S(z)g(z)\Vert_{\cal K}
\leq \Vert S(z)g(z)\Vert_{\cal K},\quad z\in\Bbb{C}_{+}.
\]
Obviously, linear submanifolds ${\cal N}$ of ${\cal N}_{+}$ 
which are invariant w.r.t. multiplication
with $e^{itz}$ yield linear submanifolds ${\cal M}:=S{\cal N}$ which are invariant
w.r.t. the semigroup $T_{+}(\cdot)$:

\vspace{3mm}

LEMMA 2. {\em Let the linear submanifold} ${\cal N}\subseteq{\cal N}_{+}$ {\em be
invariant w.r.t. multiplication with} $e^{itz}$. {\em Then
the linear manifold} ${\cal M}:=S{\cal N}$ {\em is invariant w.r.t. the
semigroup} $T_{+}(\cdot)$.

\vspace{3mm}

Proof. Let $f\in{\cal M}$. Then $(T_{+}(t)f)(\lambda)=e^{it\lambda}f(\lambda)$
and there is a $g\in{\cal N}$ satisfying (9). Then
\[
e^{itz}f(z)=e^{itz}S(z)g(z)=S(z)\{e^{itz}g(z)\},\quad z\in\Bbb{C}_{+},
\]
and the function $z\rightarrow e^{itz}g(z)$ is from ${\cal N}$,
i.e. $T_{+}(t)f\in{\cal M}$ for all $t\geq 0.\quad\Box$

\vspace{3mm}

For the orthogonal complement
${\cal H}^{2}\ominus{\cal M}$ one obtains

\vspace{3mm}

LEMMA 3. {\em The subspace} ${\cal H}^{2}_{+}\ominus{\cal M}$ {\em is invariant
w.r.t. the characteristic semigroup} $C_{+}(\cdot)$.

\vspace{3mm}

Proof. Let $g\in{\cal M}$ and $f\in{\cal H}^{2}_{+}\ominus{\cal M}$. Then
\[
(C_{+}(t)f,g)=(Q_{+}e^{-itM}f,g)=(f,e^{itM}g)
\]
and $e^{itM}g\in{\cal M}$, hence $(f,e^{itM}g)=0$ for all $g\in{\cal M}$
and $C_{+}(t)f\in{\cal H}^{2}_{+}\ominus{\cal M}$ for all $t\geq 0.\quad\Box$

\vspace{3mm}

This means that the restriction of the characteristic semigroup to invariant
subspaces ${\cal T}:={\cal H}^{2}_{+}\ominus{\cal M}$ is again a strongly
continuous contractive semigroup. We denote the generator of these restrictions
by $B_{+}$,
\[
C_{+}(t)\restriction{\cal T}=:e^{-itB_{+}},\quad t\geq 0.
\]
Note that Proposition 2(v) implies
\[
\mbox{s-lim}_{t\rightarrow\infty}\,e^{-itB_{+}}=0.
\]
The generator $B_{+}$ depends on the scattering operator, $B_{+}=B_{+}(S)$ and,
of course, on the choice of the subspace ${\cal T}$ which we call an {\em admissible}
subspace. 
In the following section the spectrum of $B_{+}$ is calculated in selected cases under
additional assumptions for $S$.

\section{Results}

The first additional conditions for $S(\cdot)$, assumed throughout in the following, are 
\begin{itemize}
\item[(iii)]
The scattering matrix $S(\cdot)$ has no mutually complex conjugated poles
in \\$\Bbb{C}\setminus(-\infty,0]$
including the rims $\Bbb{R}_{-}\pm i0$), i.e. if $\zeta\in\Bbb{C}_{-}$ is a pole of $S(\cdot)$
then it is holomorphic at $\overline{\zeta}\in\Bbb{C}_{+}$.
\item[(iv)]
$S(\cdot)$ has at least one pole in the lower half plane.
\end{itemize}
The conditions (iii) and (iv) 
ensure that admissible subspaces ${\cal T}$ are not $\{0\}$.

\vspace{3mm}

LEMMA 4. {\em Let} $\zeta\in\Bbb{C}_{-}$ {\em be a pole of} $S(\cdot)$. {\em Then}
$\ker\,S(\overline{\zeta})^{\ast}\supset\{0\}$. Let $k\in\ker\,
S(\overline{\zeta})^{\ast}$, {\em i.e.} $S(\overline{\zeta})^{\ast}k=0$. 
{\em Further let} ${\cal T}:={\cal H}^{2}_{+}\ominus{\cal M}$ 
{\em be an admissible subspace. Then}
$e\in{\cal T}$ {\em where}
\[
e(\lambda):=\frac{k}{\lambda-\zeta}.
\]

\vspace{3mm}

Proof. Obviously $e\in{\cal H}^{2}_{+}$.
Let $g(z):=S(z)f(z)$ where $g\in{\cal M},\,f\in{\cal N}$. Then
\begin{eqnarray*}
(e,g)=\int_{-\infty}^{\infty}\left(\frac{k}{\lambda-\zeta},g(\lambda+i0)\right)_{\cal K}
d\lambda &=&
\int_{-\infty}^{\infty}\frac{1}{\lambda-\overline{\zeta}}(k,g(\lambda+i0))_{\cal K}
d\lambda\\
&=& 2\pi i(k,g(\overline{\zeta}))_{\cal K}\\
&=& 2\pi i(k,S(\overline{\zeta})f(\overline{\zeta}))_{\cal K}\\
&=& 2\pi i(S(\overline{\zeta})^{\ast}k,f(\overline{\zeta}))_{\cal K}=0
\end{eqnarray*}
for all $f\in{\cal N}$, i.e. for all $g\in{\cal M}$, i.e. $e\in{\cal T}.\quad\Box$

\vspace{3mm}

Note that so far the case ${\cal M}_{+}=\{0\}$ is not excluded. In this case there is no
proper restriction of the characteristic semigroup, i.e. ${\cal T}={\cal H}^{2}_{+}$.

\vspace{3mm}

As mentioned in the introduction, the cases
\begin{equation}
S(\lambda+i0)=S(\lambda-i0),\quad \lambda<0
\end{equation}
and
$S(\lambda+i0)\neq S(\lambda-i0)$ for $\lambda<0$ are completely different. For example,
in the first case, considered in Subsection 5.1,
there are necessarily no poles on the negative half line. 
In the second case
poles on the rims $\Bbb{R}_{-}\pm i0$ are not excluded. Especially these cases are
of interest because poles on the upper rim indicate the existence of
eigenvalues of $H$ (see e.g. examples in potential scattering). This case
is presented in Subsection 5.2, where we assume that there are
no poles on the upper half plane but finitely many poles on the rims 
$\Bbb{R}_{-}\pm i0$.

\subsection{The case $S(\lambda+i0)=S(\lambda-i0)$ for $\lambda<0$}

In this case we use (ii)(iii), and (iv). Note that in this case $S(\cdot)$ is well-defined
and meromorphic on the unique sheet $\Bbb{C}\setminus\{0\}$. Thus we
can put $S(\lambda\pm i0)=:S(\lambda)$ also for $\lambda<0$.
It is unitary also on
the negative half line, i.e. without poles there. That is, in this case
\[
\Bbb{R}\ni\lambda\rightarrow S(\lambda)
\]
is a unitary operator function on ${\cal K}$ on the whole real line defining a unitary
operator on ${\cal H}$ which we denote, without ambiguity, by $S$. Further we assume 

\begin{itemize}
\item[(v)]
There are at most finitely many poles of $S(\cdot)$ in the upper half plane, $S(\cdot)$
is bounded at $z=0$ and there is
a constant $R>0$ such that the poles ly inside the semi-circle $\{z\in\Bbb{C}_{+}:
\vert z\vert<R\}$ and
\[
\sup_{\{z\in\Bbb{C}_{+}:\vert z\vert>R\}}\Vert S(z)\Vert=:K<\infty.
\]
\end{itemize}
Note that (v) implies that in the present case (11) $S(\cdot)$ is holomorphic at $z=0$, too.
In this case one obtains ${\cal M}_{+}\supset\{0\}$. We denote these poles in $\Bbb{C}_{+}$
by $\xi_{1},\xi_{2},...,\xi_{r}$. The order of the pole $\xi_{j}$ is $g_{j}$.

\vspace{3mm}

LEMMA 5. {\em Let} ${\cal N}_{\xi,g}\subset{\cal H}^{2}_{+}$ {\em be the subspace of
Lemma 1. Then} $S{\cal N}_{\xi,g}\subset{\cal H}^{2}_{+}$, {\em i.e.}
\begin{equation}
S{\cal N}_{\xi,g}\subseteq{\cal M}_{+}.
\end{equation}

\vspace{3mm}

Proof. Let $p$ be the polynomial $p(\lambda):=\Pi_{j=1}^{r}(\lambda-\xi_{j})^{g_{j}}$.
Then $z\rightarrow p(z)S(z)$ is holomorphic on $\Bbb{C}_{+}$ and if
$f\in{\cal N}_{\xi,g}$ then $f(z)=p(z)g(z)$ where $g\in{\cal H}^{2}_{+}$. Then an easy
calculation shows that $z\rightarrow S(z)f(z)=p(z)S(z)g(z)$ is from ${\cal H}^{2}_{+}.
\quad\Box$

\vspace{3mm}

Note that in this case
the linear manifold $S{\cal N}_{\xi,g}$ in (12) is a subspace. Note further that the subspace
${\cal N}_{\xi,g}$ is invariant w.r.t. the multiplication with the function
$z\rightarrow e^{itz}$ because this function does not vanish everywhere. In the following
we choose the admissible subspace ${\cal T}:={\cal H}^{2}_{+}\ominus S{\cal N}_{\xi,g}$.

\vspace{3mm}

THEOREM 2. {\em Assume conditions} (iii),(iv),(v) {\em and} (11). {\em Then}
$\{0\}\subset{\cal T}\subset{\cal H}^{2}_{+}$. {\em The spectrum}
$\mbox{spec}\,B_{+}\subseteq\Bbb{C}_{-}\cup\Bbb{R}$
{\em of the generator} $B_{+}$ {\em of the restriction of the characteristic
semigroup onto} ${\cal T}$ {\em is described as follows}:
\begin{itemize}
\item[(i)]
{\em If} $\zeta\in\Bbb{C}_{-}$ {\em then} $\zeta\in\mbox{res}\,B_{+}$ {\em iff}
$S(\overline{\zeta})^{\ast}$ {\em is invertible, i.e. if}
$(S(\overline{\zeta})^{\ast})^{-1}=S(\zeta)$ {\em exists, i.e. if} $S(\cdot)$
{\em is holomorphic at} $\zeta$.
\item[(ii)]
{\em If} $\zeta\in\Bbb{C}_{-}$ {\em then} $\zeta$ {\em is an eigenvalue of}
$B_{+}$ {\em iff} $\ker\,S(\overline{\zeta})^{\ast}\supset\{0\}$, {\em i.e. if}
$\zeta$ {\em is a pole of} $S(\cdot)$, {\em i.e. if} $\zeta$ {\em is a resonance. The
corresponding eigenvectors are given by}
\[
e_{\zeta,k}(\lambda):=\frac{k}{\lambda-\zeta},\quad k\in\ker\,S(\overline{\zeta})^{\ast}.
\]
\item[(iii)]
{\em If} $\lambda\in\Bbb{R}$ {\em then} $\lambda\in\mbox{res}\,B_{+}$.
\end{itemize}

That is, the eigenvalue spectrum of $B_{+}$ coincides with the set ${\cal R}$ of all
resonances and ${\cal R}=\mbox{spec}\,B_{+}$.

\vspace{3mm}

Proof. First we prove (ii). Let $\zeta\in\Bbb{C}_{-}$ be an eigenvalue of $B_{+}$ and
$e$ a corresponding eigenvector. Now
$\zeta$ is necessarily also an eigenvalue of $C_{+}$, the generator of the
characteristic semigroup.
Thus there is a vector $k\in{\cal K}$ such that
\[
e(\lambda)=\frac{k}{\lambda-\zeta}.
\]
Since $e\in{\cal T}$ this means that it is orthogonal w.r.t. $S{\cal N}_{\xi,g}$.
That is, we obtain for all $g\in{\cal N}_{\xi,g}$
\begin{eqnarray*}
0=\int_{-\infty}^{\infty}\left(\frac{k}{\lambda-\zeta},S(\lambda)g(\lambda)\right)_{\cal K}
d\lambda &=&
\int_{-\infty}^{\infty}\frac{1}{\lambda-\overline{\zeta}}
(k,S(\lambda)g(\lambda))d\lambda\\
&=& 2\pi i(k,S(\overline{\zeta})g(\overline{\zeta}))_{\cal K}\\
&=& 2\pi i(S(\overline{\zeta})^{\ast}k,g(\overline{\zeta}))_{\cal K}.
\end{eqnarray*}
The vectors $g(\overline{\zeta})$ exhaust ${\cal K}$, because, for example, the
functions
\[
g(z)=\frac{p(z)}{(z+i)^{g+1}}k,\quad k\in{\cal K},
\]
are from ${\cal N}_{\xi,g}$. Therefore $S(\overline{\zeta})^{\ast}k=0$ follows.
The converse is obvious.

(i) Let $\zeta\in\Bbb{C}_{-}$ and assume that $S(\overline{\zeta})^{-1}$ exists. Then
we have to prove that $\zeta\in\mbox{res}\,B_{+}$. 
Since $\ker\,S(\overline{\zeta})^{\ast}=\{0\}$ hence $\zeta$ is not an eigenvalue
of $B_{+}$ it follows that $(B_{+}-\zeta\EINS)^{-1}$ exists. Therefore it is
sufficient to show that $\mbox{ima}\,(B_{+}-\zeta\EINS)={\cal T}$. Let $g\in{\cal T}$.
Then we have to construct a function $f\in\mbox{dom}\,B_{+}$ with
$(B_{+}-\zeta\EINS)f=g$. That is $f\in{\cal H}^{2}_{+}\ominus S{\cal N}_{\xi,g}$
and $f\in\mbox{dom}\,C_{+}$ and this means that we have to construct a vector
$k_{0}\in{\cal K}$ such that the function $\lambda\rightarrow\lambda f(\lambda)+k_{0}$
is from ${\cal H}^{2}_{+}$ and orthogonal to the subspace $S{\cal N}_{\xi,g}$.
Then $B_{+}f(\lambda)=\lambda f(\lambda)+k_{0}$ and we have to prove
$\lambda f(\lambda)+k_{0}-\zeta f(\lambda)=g(\lambda)$ or
\[
f(\lambda)=\frac{g(\lambda)-k_{0}}{\lambda-\zeta}.
\]
Obviously, in any case one has $f\in{\cal H}^{2}_{+}$. Now $g$ is orthogonal to
$S{\cal N}_{\xi,g}$ or $S^{\ast}g$ is orthogonal to ${\cal N}_{\xi,g}$. This means
$S^{\ast}g=Q_{-}(S^{\ast}g)+P(S^{\ast}g)$, where $P$ denotes the projection onto
${\cal H}^{2}_{+}\ominus{\cal N}_{\xi,g}$. Then
\begin{equation}
S(\lambda)^{\ast}f(\lambda)=\frac{S(\lambda)^{\ast}g(\lambda)-S(\lambda)^{\ast}k_{0}}
{\lambda-\zeta}=\frac{Q_{-}(S^{\ast}g)(\lambda)-S(\lambda)^{\ast}k_{0}}
{\lambda-\zeta}+\frac{P(S^{\ast}g)(\lambda)}{\lambda-\zeta}.
\end{equation}
We put $h(z):=Q_{-}(S^{\ast}g)(z)$. This function is holomorphic on the lower half plane
hence $h(\zeta)$ exists and we fix $k_{0}$ by 
$k_{0}:=(S(\overline{\zeta})^{\ast})^{-1}h(\zeta)$. Then 
$S(\overline{\zeta})^{\ast}k_{0}=h(\zeta)$. Thus the first term on the right hand side
of (13) is holomorphic at $z=\zeta$ hence it is from ${\cal H}^{2}_{-}$ and
a fortiori orthogonal to ${\cal N}_{\xi,g}$. Concerning the second term note that
the function $u(z):=\frac{1}{z-\zeta}$ does not vanish on $\Bbb{C}_{+}$, hence
${\cal N}_{\xi,g}$ is invariant w.r.t. multiplication with $u$, i.e.
$u{\cal N}_{\xi,g}\subseteq{\cal N}_{\xi,g}$. This implies 
$u({\cal H}^{2}_{+}\ominus{\cal N}_{\xi,g})\subseteq{\cal H}^{2}_{+}\ominus{\cal N}_{\xi,g}$.
Thus the second term is orthogonal to ${\cal N}_{\xi,g}$, too. Therefore we obtain
from (13) that $f\in{\cal H}^{2}_{+}\ominus{\cal N}_{\xi,g}$ and $\zeta\in\mbox{res}\,B_{+}$.
Conversely, let $\zeta\in\mbox{res}\,B_{+}$. One has to show that
$(S(\overline{\zeta})^{\ast})^{-1}$ exists. Assume, on the contrary, 
that $S(\overline{\zeta})^{\ast}$ is not invertible. Then 
$\ker\,S(\overline{\zeta})^{\ast}\supset\{0\}$ and therefore, according to (ii), $\zeta$
is an eigenvalue of $B_{+}$, a contradiction.

(iii) Let $\mu_{0}\in\Bbb{R}$. One has to show $\mu_{0}\in\mbox{res}\,B_{+}.
\;\mu_{0}$ is a point of the spectrum of the characteristic semigroup $C_{+}(\cdot)$ but
it is not an eigenvalue hence, a fortiori, not an eigenvalue of $B_{+}$, too. Thus
$(B_{+}-\mu_{0}\EINS)^{-1}$ exists. The assertion is $\mbox{ima}\,(B_{+}-\mu_{0} 
\EINS)={\cal T}$. Let $g\in{\cal T}$. 
Then $S^{\ast}g$ is orthogonal to ${\cal N}_{\xi,g}$, i.e.
\begin{equation}
S(\lambda)^{\ast}g(\lambda)=Q_{-}(S^{\ast}g)(\lambda)+P(S^{\ast}g)(\lambda)
\end{equation}
Recall Lemma 1(ii). Since there are only finitely many poles $\xi_{j}$ there is an
$\epsilon>0$ such that the functions
\begin{equation}
z\rightarrow\frac{k}{(z-\overline{\xi_{j}})^{m_{j}}}
\end{equation}
are holomorphic even on $\Bbb{C}_{+}-i\epsilon.$ Therefore they can be considered as
elements of the Hardy space ${\cal H}^{2}_{+}$ where the "real line" is shifted to
$\Bbb{R}-i\epsilon$ and the closure of the span of the functions (15) is a subspace
of this Hardy space. This implies that the second term of (14) is holomorphic on 
$\Bbb{R}\cup\Bbb{C}_{+}$. From (14) we obtain
\begin{equation}
g(\lambda)-P(S^{\ast}g)(\lambda)=S(\lambda)Q_{-}(S^{\ast}g)(\lambda).
\end{equation}
The left hand side of (16) is holomorphic in the "upper neighborhood" of $\mu_{0}$ and
the right hand side is holomorphic in its "lower neighborhood". Then, according to
"Schwarzsches Spiegelungsprinzip" it follows that $g$ is holomorphic at $\mu_{0}$. The
next step corresponds to that in (i): We have to choose a $k_{0}\in{\cal K}$ such that
the function $f(\lambda):=\frac{g(\lambda)-k_{0}}{\lambda-\mu_{0}}$ is an element
of ${\cal T}$. We put $k_{0}:=g(\mu_{0})$. Then
\begin{equation}
\Bbb{R}\ni\lambda\rightarrow f(\lambda):=
\frac{g(\lambda)-g(\mu_{0})}{\lambda-\mu_{0}}
\end{equation}
is holomorphic at $\lambda=\mu_{0}$, too. First, $f\in{\cal H}^{2}_{+}$. We have to
show that $S^{\ast}f$ is orthogonal to ${\cal N}_{\xi,g}$. For this reason
we put
\[
f_{\epsilon}(\lambda):=\frac{g(\lambda)-g(\mu_{0})}{\lambda-(\mu_{0}-i\epsilon)},
\quad \epsilon>0.
\]
Obviously, $f_{\epsilon}\in{\cal H}^{2}_{+},\;
\mbox{s-lim}_{\epsilon\rightarrow +0}f_{\epsilon}=f$ and also 
$\mbox{s-lim}_{\epsilon\rightarrow +0}S^{\ast}f_{\epsilon}=S^{\ast}f$. We calculate
with an arbitrary $u\in{\cal N}_{\xi,g}$
\begin{eqnarray*}
(S^{\ast}f_{\epsilon},u)&=&
\int_{-\infty}^{\infty}\left(\frac{Q_{-}(S^{\ast}g)(\lambda)+P(S^{\ast}g)(\lambda)
-S(\lambda)^{\ast}g(\mu_{0})}{\lambda-(\mu_{0}-i\epsilon)},u(\lambda)\right)_{\cal K}d\lambda\\
&=&
\int_{-\infty}^{\infty}\frac{1}{\lambda-(\mu_{0}+i\epsilon)}
(Q_{-}(S^{\ast}g)(\lambda),u(\lambda))_{\cal K}d\lambda+
\int_{-\infty}^{\infty}\frac{1}{\lambda-(\mu_{0}+i\epsilon)}(P(S^{\ast}g)(\lambda),
u(\lambda))_{\cal K}d\lambda\\
\end{eqnarray*}
\[
-\int_{-\infty}^{\infty}\frac{1}{\lambda-(\mu_{0}+i\epsilon)}
(S(\lambda)^{\ast}g(\mu_{0}),u(\lambda))_{\cal K}d\lambda.
\]
For the first and the third term on the right hand side we obtain
\begin{equation}
\int_{-\infty}^{\infty}\frac{1}{\lambda-(\mu_{0}+i\epsilon)}(Q_{-}(S^{\ast}g)(\lambda),
u(\lambda))_{\cal K}d\lambda=
2\pi i(Q_{-}(S^{\ast}g)(\mu_{0}-i\epsilon),u(\mu_{0}+i\epsilon))
\end{equation}
and
\begin{equation}
\int_{-\infty}^{\infty}\frac{1}{\lambda-(\mu_{0}+i\epsilon)}
(S(\lambda)^{\ast}g(\mu_{0}),u(\lambda))d\lambda=2\pi i(g(\mu_{0}),S(\mu_{0}+i\epsilon)
u(\mu_{0}+i\epsilon))_{\cal K}.
\end{equation}
For the second term note that
\[
\int_{-\infty}^{\infty}\frac{1}{\lambda-(\mu_{0}+i\epsilon)}\left(
\frac{k}{(\lambda-\overline{\xi})^{r}},u(\lambda)\right)d\lambda=
2\pi i\left(\frac{k}{(\mu_{0}-i\epsilon-\overline{\xi})^{r}},u(\mu_{0}+i\epsilon)\right)
\]
where $\xi$ denotes one of the poles of $S(\lambda)$ in the upper half plane. Note
further that these functions $\lambda\rightarrow\frac{k}{(\lambda-\overline{\xi})^{r}}$
generate the subspace ${\cal H}^{2}_{+}\ominus{\cal N}_{\xi,g}$. Thus for this term we
obtain
\begin{equation}
\int_{-\infty}^{\infty}\frac{1}{\lambda-(\mu_{0}+i\epsilon)}
(P(S^{\ast}g)(\lambda),u(\lambda))d\lambda=2\pi i(P(S^{\ast}g)(\mu_{0}-i\epsilon),
u(\mu_{0}+i\epsilon)).
\end{equation}
Putting together the results (18),(19),(20) we have
\[
(S^{\ast}f_{\epsilon},u)=
2\pi i(S(\mu_{0}-i\epsilon)^{\ast}g(\mu_{0}-i\epsilon)-S(\mu_{0}+i\epsilon)^{\ast}g(\mu_{0}),
u(\mu_{0}+i\epsilon))).
\]
Taking the limit $\epsilon\rightarrow+0$ we obtain $(S^{\ast}f,u)=0$ for all 
$u\in{\cal N}_{\xi,g}$, i.e. $f\in{\cal T}.\quad\Box$
 
\vspace{3mm}

EXAMPLE 1. A simple example for this case is given by the asymptotic complete scattering
system $\{H,M\}$ on ${\cal H}:=L^{2}(\Bbb{R})$ where
$H:=M+(h,\cdot)h$ and $h(\lambda):=(\pi(\lambda^{2}+1))^{-1/2}$, i.e. 
$\Vert h\Vert=1$. Then
\[
S(\lambda)=\frac{\lambda+i}{\lambda-i}\cdot\frac{\lambda-1-i}{\lambda-1+i}
\]
with the poles $\xi:=i$ and $\zeta:=1-i$. (Of course there is $H_{+}$ on $L^{2}(\Bbb{R}_{+})$
such that $\{H_{+},M_{+}\}$ is an asymptotic complete scattering system with the 
same scattering matrix, restricted to $\Bbb{R}_{+}$). In this case ${\cal N}_{\xi}$
is the subspace of all Hardy functions $f$ with $f(i)=0$ and 
$S{\cal N}_{\xi}={\cal N}_{1+i}$ the corresponding subspace where $f(1+i)=0$.
Further ${\cal T}=\Bbb{C}e_{\zeta}$ where $e_{\zeta}(\lambda)=\frac{1}{\lambda-\zeta}$,
i.e. ${\cal T}$ is one-dimensional and the semigroup $C_{+}(t)\restriction{\cal T}$
acts as multiplication by $e^{-it\zeta}$. The vector $k_{0}$ for $e_{\zeta}$
is simply $-1$.

\subsection{The case $S(\lambda+i0)\neq S(\lambda-i0)$ for $\lambda<0$}

In many quantum mechanical scattering systems of physical interest the scattering
matrix is not a unique analytic function on $\Bbb{C}$ but its Riemann surface
has several sheets. For example, in the case of potential scattering it is often
that of $z^{1/2}$, i.e. there are two sheets, where the physical scattering matrix
lives on $\Bbb{R}_{+}$ of the first sheet and its inverse on $\Bbb{R}_{+}$ of the
second sheet, i.e. for every $z\neq 0$ the values of $S(\cdot)$ on the two sheets
are mutually inverse. Further it occurs often that there are no poles 
in the upper half plane of the first sheet but there are poles of $S(\cdot)$
on the upper rim $\Bbb{R}_{-}+i0$ due to
the existence of eigenvalues of $H$ and poles (resonances) in
the lower half plane of this sheet. 

Therefore in this section we focus on the property
\begin{equation}
S(\lambda+i0)\neq S(\lambda-i0),\quad \lambda<0
\end{equation}
and omit the complication of poles in the upper half plane but assume the existence
of (finitely many) poles on $\Bbb{R}_{-}+i0$. (21) implies
\[
S(\lambda-i0)^{-1}=S(\lambda+i0)^{\ast},\quad \lambda<0,
\]
i.e. in this case the scattering matrix cannot be unitarily extended onto the
negative half line. We assume, as before, (ii),(iii),(iv).
(v) is replaced by
\begin{itemize}
\item[(v')]
$S(\cdot)$ is holomorphic on the upper half plane, there are finitely many poles
on the upper rim $\Bbb{R}_{-}+i0,\;S(\cdot)$ is bounded at $z=0$ and there is a constant
$R>0$ such that the poles 
ly inside the semi-circle $\{z\in\Bbb{C}_{+}:\vert z\vert<R\}$ and
\[
\sup_{\{z\in\Bbb{C}_{+}:\vert z\vert>R\}}\Vert S(z)\Vert:=K<\infty.
\]
\end{itemize}
We choose the admissible subspace ${\cal T}:={\cal H}^{2}_{+}\ominus{\cal M}_{+}$. 
Also in this case
${\cal M}_{+}=S{\cal N}_{+}\supset\{0\}$.

\vspace{3mm}

LEMMA 6. {\em Let} $p$ {\em be the polynomial} 
$p(\lambda):=\Pi_{j=1}^{r}(\lambda+a_{j})^{g_{j}},\;a_{j}>0$, {\em where} $-a_{1},-a_{2},...
-a_{r}$ {\em are the poles of} $S(\cdot+i0)$ {\em on} $\Bbb{R}_{-}+i0$ {\em and}
$g_{j}\in\Bbb{N}$ {\em denotes the order of the pole} $-a_{j}$. {\em Then all
functions} $v(\cdot)$ {of the form}
\begin{equation}
v(\lambda):=\frac{p(\lambda)}{(\lambda+i)^{g}}w(\lambda),
\quad w\in{\cal H}^{2}_{+},
\end{equation}
{\em where} $g$ {\em is the order of the polynomial} $p$,
{\em are elements of} ${\cal N}_{+}$.

\vspace{3mm}

Proof. Obvious because of $v\in{\cal H}^{2}_{+},\; z\rightarrow p(z)S(z)$ is
holomorphic on the upper half plane including the rim $\Bbb{R}_{-}+i0$, bounded
at $z=0$, hence
\begin{equation}
\sup_{\{z\in\Bbb{C}_{+}:\vert z\vert\leq R\}}\Vert p(z)S(z)\Vert<\infty
\end{equation}
and
\[
\Vert S(z)v(z)\Vert\leq K\Vert v(z)\Vert,\quad \vert z\vert>R.\quad\Box
\]

\vspace{3mm}

In this context the spectrum of the generator $B_{+}$ of the restriction of
the characteristic semigroup onto ${\cal T}$ coincides again with the set of
all resonances like in Theorem 2. 

\vspace{3mm}

THEOREM 3. {\em Assume conditions} (iii),(iv),(v') {\em and} (21). {\em Then}
${\cal T}:={\cal H}^{2}_{+}\ominus{\cal M}_{+}$ {\em is admissible and}
$\{0\}\subset{\cal T}\subset{\cal H}^{2}_{+}$. {\em The spectrum}
$\mbox{spec}\,B_{+}\subset\Bbb{C}_{-}\cup\Bbb{R}$ {\em of} $B_{+}$
{\em is described as follows}:
\begin{itemize}
\item[(i)]
{\em If} $\zeta\in\Bbb{C}_{-}$ {\em then} $\zeta$ {\em is an eigenvalue of}
$B_{+}$ {\em iff} $\ker\,S(\overline{\zeta})^{\ast}\supset\{0\}$, {\em i.e. if}
$\zeta$ {\em is a pole of} $S(\cdot)$, {\em i.e. if} $\zeta$ {\em is a resonance.
The corresponding eigenvectors are given by}
\[
e_{\zeta,k}(\lambda):=\frac{k}{\lambda-\zeta},\quad k\in\ker\,S(\overline{\zeta}^{\ast}).
\]
\item[(ii)]
{\em If} $\zeta\in\Bbb{C}_{-}$ {\em then} $\zeta\in\mbox{res}\,B_{+}$ {\em iff}
$S(\overline{\zeta})^{\ast}$ {\em is invertible, i.e. if}
$(S(\overline{\zeta})^{\ast})^{-1}=S(\zeta)$ {\em exists, i.e. if} $S(\cdot)$
{\em is holomorphic at} $\zeta$.
\item[(iii)]
{\em If} $\mu\in\Bbb{R}$ {\em then} $\mu\in\mbox{res}\,B_{+}$
\end{itemize}

\vspace{3mm}

Proof. The proof of (i) is similar as that of (ii) in Theorem 2. Now $e\in{\cal T}$
means that $e$ is orthogonal w.r.t. $S{\cal N}_{+}$. That is, for all $v\in{\cal N}_{+}$
we obtain again
\[
0=\int_{-\infty}^{\infty}\left(\frac{k}{\lambda-\zeta},S(\lambda)v(\lambda)\right)_{\cal K}
d\lambda=2\pi i(S(\overline{\zeta})^{\ast}k,v(\overline{\zeta}))_{\cal K},
\]
but again the vectors $v(\overline{\zeta})$ exhaust ${\cal K}$, e.g. choose
$w(\lambda):=\frac{k}{\lambda+i}$ in (22).

(ii) Let $\zeta\in\Bbb{C}_{-}$ and assume that $S(\overline{\zeta})^{-1}$ exists.
Then the assertion is $\zeta\in\mbox{res}\,B_{+}$. The first arguments follow
the lines of the proof of (i) in Theorem 2. Then, again for $g\in{\cal T}$ one has
to construct $k_{0}\in{\cal K}$ such that the function
\begin{equation}
f(\lambda):=\frac{g(\lambda)-k_{0}}{\lambda-\zeta}
\end{equation}
is an element of ${\cal T}$. According to (22) we have for all $w\in{\cal H}^{2}_{+}$ 
\begin{equation}
\int_{-\infty}^{\infty}\left(g(\lambda),S(\lambda+i0)\frac{p(\lambda)}{(\lambda+i)^{g}}
w(\lambda)\right)_{\cal K}d\lambda=
\int_{-\infty}^{\infty}\left(\frac{p(\lambda)}{(\lambda-i)^{g})}
S(\lambda+i0)^{\ast}g(\lambda),w(\lambda)\right)_{\cal K}d\lambda=0,
\end{equation}
where the function
\begin{equation}
\Bbb{R}\ni\lambda\rightarrow h_{-}(\lambda):=
\frac{p(\lambda)}{(\lambda-i)^{g}}S(\lambda+i0)^{\ast}g(\lambda)
\end{equation}
is an element of $L^{2}(\Bbb{R},{\cal K})$, because, according to (v') and (23),
we have
\[
\sup_{\lambda\in\Bbb{R}}\Vert\frac{p(\lambda)}{(\lambda-i)^{g}}S(\lambda+i0)^{\ast}
\Vert<\infty.
\]
Now (25) implies $h_{-}\in{\cal H}^{2}_{-}$. This gives, note that
$S(\lambda-i0)=S(\lambda+i0)=S(\lambda)$ for $\lambda>0$,
\begin{equation}
g(\lambda)=\frac{(\lambda-i)^{g}}{p(\lambda)}S(\lambda-i0)h_{-}(\lambda).
\end{equation}
The right hand side of (27) is meromorphic on $\Bbb{C}_{-}$, the left hand side is
holomorphic on $\Bbb{C}_{+}$.
According to "Schwarzsches Spiegelungsprinzip" this means that $g$ is meromorphic
on $\Bbb{R}\setminus\{0\}$ with poles at most at the poles of $S(\cdot-i0)$ and
at the zeros of $p(\cdot)$. But $g\in{\cal H}^{2}_{+}$ and poles are not
locally square integrable hence $g$ is holomorphic on $\Bbb{R}\setminus\{0\}$. Then (26)
and (v') imply that $g$ is holomorphic at $z=0$, too. 
Therefore, $g(\cdot)$ is meromorphic on $\Bbb{C}$, possible poles are the poles
of $S(\cdot)$ in $\Bbb{C}_{-}$.

Because of (24) we have
$f\in{\cal H}^{2}_{+}$. It is required that $f\in{\cal T}$. This means
\[
\int_{-\infty}^{\infty}(f(\lambda),u(\lambda+i0))_{\cal K}d\lambda=0
\]
for all $u\in{\cal M}_{+}=S{\cal N}_{+}$, i.e. $u(z)=S(z)v(z),\;z\in\Bbb{C}_{+},\;
v\in{\cal N}_{+}$ or
\begin{equation}
\int_{-\infty}^{\infty}\frac{1}{\lambda-\overline{\zeta}}\left(g(\lambda),
u(\lambda+i0)\right)_{\cal K}d\lambda=
\int_{-\infty}^{\infty}\frac{1}{\lambda-\overline{\zeta}}(k_{0},u(\lambda+i0))_{\cal K}d\lambda.
\end{equation}
In particular (28) holds for all $u:=Sv$ where
 $v\in{\cal N}_{+}$ is of the form (22). Inserting these $u$
into the right hand side of (28) one obtains the term
\begin{equation}
2\pi i\left(k_{0}, S(\overline{\zeta})\frac{p(\overline{\zeta})}{(\overline{\zeta}+i)^{g}}
w(\overline{\zeta})\right)_{\cal K},
\end{equation}
and for the left hand side the term
\begin{equation}
\int_{-\infty}^{\infty}\frac{1}{\lambda-\overline{\zeta}}\left(
\frac{p(\lambda)}{(\lambda-i)^{g}}S(\lambda+i0)^{\ast}g(\lambda),w(\lambda)\right)_{\cal K}
d\lambda
=\int_{-\infty}^{\infty}\frac{1}{\lambda-\overline{\zeta}}(h_{-}(\lambda),
w(\lambda))_{\cal K}d\lambda
\end{equation}
\[
=2\pi i(h_{-}(\zeta),w(\overline{\zeta}))_{\cal K}.
\]
Since the values $w(\overline{\zeta})$ exhaust ${\cal K}$ (28) is satisfied iff
\begin{equation}
h_{-}(\zeta)=\frac{p(\zeta)}{(\zeta-i)^{g}}S(\overline{\zeta})^{\ast}k_{0}.
\end{equation}
Therefore, putting
\[
k_{0}:=\frac{(\zeta-i)^{g}}{p(\zeta)}(S(\overline{\zeta})^{\ast})^{-1}h_{-}(\zeta)=g(\zeta),
\]
(28) is satisfied for all $v\in{\cal N}_{+}$. Finally we show that 
with this $k_{0}$ equation
(28) is also true for all $u\in{\cal M}_{+}$. For the right hand side of (28)
we obtain the expression
\begin{equation}
2\pi i(k_{0},u(\overline{\zeta}))_{\cal K}=
2\pi i(k_{0},S(\overline{\zeta})v(\overline{\zeta}))_{\cal K}.
\end{equation}
The function $\lambda\rightarrow (g(\lambda),u(\lambda+i0))_{\cal K}$ of the integrand on the
left hand side is a scalar $L^{2}$-function on $\Bbb{R}$ which is analytically
continuable onto $\Bbb{C}_{+}$ by
$\Bbb{C}_{+}\ni z\rightarrow (g(\overline{z}),S(z)v(z))_{\cal K}.$
Note that
$g(\cdot)$ has poles on $\Bbb{C}_{-}$. However 
\[
(g(\overline{z}),S(z)v(z))_{\cal K}=(S(z)^{\ast}g(\overline{z}),v(z))_{\cal K}=
(S(\overline{z})^{-1}g(\overline{z}),v(z))_{\cal K}
\]
and, according to (27), we have
\begin{equation}
S(z)^{-1}g(z)=\frac{(z-i)^{g}}{p(z)}h_{-}(z),\quad z\in\Bbb{C}_{-},
\end{equation}
which is holomorphic there. Therefore the left hand side of (28)
is given by
\[
2\pi i\left(\frac{(\zeta-i)^{g}}{p(\zeta)}h_{-}(\zeta),v(\overline{\zeta})\right)_{\cal K},
\]
hence because of (31) it coincides with (32) and
$(B_{+}-\zeta\EINS)f=g$ is true, i.e. $\zeta\in\mbox{res}\,B_{+}.$

(iii) Let $\mu_{0}\in\Bbb{R}$. The assertion is $\mu_{0}\in\mbox{res}\,B_{+}$.
As in the proof of (iii) in Theorem 2 to every $g\in{\cal T}$ we have to construct
$k_{0}\in{\cal K}$ such that the function
\[
f(\lambda):=\frac{g(\lambda-k_{0}}{\lambda-\mu_{0}}
\]
is an element of ${\cal T}$, i.e. orthogonal w.r.t. ${\cal M}_{+}$. According to (27)
the function $g$ is holomorphic on $\Bbb{R}$. Therefore we put $k_{0}:=g(\mu_{0})$. 
Then $f$ is holomorphic at $\mu_{0}$, too and $f\in{\cal H}^{2}_{+}.$ As before we
introduce the functions
\[
f_{\epsilon}(\lambda):=\frac{g(\lambda)-g(\mu_{0})}{\lambda-(\mu_{0}-i\epsilon)}.
\]
Then $f_{\epsilon}\in{\cal H}^{2}_{+}$ and $\mbox{s-lim}_{\epsilon\rightarrow+0}
f_{\epsilon}=f$. We calculate th integral
\[
\int_{-\infty}^{\infty}\left(\frac{g(\lambda)-g(\mu_{0})}{\lambda-(\mu_{0}-i\epsilon)},
u(\lambda+i0)\right)_{\cal K}d\lambda=
\]
\begin{equation}
\int_{-\infty}^{\infty}\frac{1}{\lambda-(\mu_{0}+i\epsilon)}
(g(\lambda),u(\lambda+i0))_{\cal K}d\lambda-
\int_{-\infty}^{\infty}\frac{1}{\lambda-(\mu_{0}+i\epsilon)}(g(\mu_{0}),u(\lambda+i0))_
{\cal K}d\lambda
\end{equation}
for all $u\in{\cal M}_{+},\; u:=Sv,\;v\in{\cal N}_{+}$. For the second term in (33) we obtain
\begin{equation}
2\pi i(g(\mu_{0}),u(\mu_{0}+i\epsilon))_{\cal K}=2\pi i(g(\mu_{0}),
S(\mu_{0}+i\epsilon)v(\mu_{0}+i\epsilon))_{\cal K},
\end{equation}
and for the first term, using once more (33),
\begin{equation}
2\pi i(S(\mu_{0}-i\epsilon)^{-1}g(\mu_{0}-i\epsilon),v(\mu_{0}+i\epsilon))_{\cal K}=
2\pi i(g(\mu_{0}-i\epsilon),S(\mu_{0}+i\epsilon)v(\mu_{0}+i\epsilon))_{\cal K}.
\end{equation}
Taking the limit $\epsilon\rightarrow+0$ the left hand side of (34) tends to
$(f,u)$ and the right hand side to $0$. Thus $f\in{\cal T}.\quad\Box$

\vspace{3mm}

EXAMPLE 2. An example for the case (21) and Theorem 3 is given by the potential
scattering with a real-valued central-symmetric potential with compact
support and zero angular momentum. In this case ${\cal K}=\Bbb{C}$ and the
(scalar) scattering matrix is given by
\[
S(E):=\frac{F(-k)}{F(k)}\quad E>0,\quad E=k^{2},\quad k>0,
\]
where $k\rightarrow F(k)$ denotes the so-called Jost function which is an entire
function of $k$ such that the Riemann surface of $S(\cdot)$ is that of
$\sqrt{E}$. In this case $S(\cdot)$ is holomorphic in the upper half plane (of the first
sheet), poles on the upper (and lower) rim $\Bbb{R}_{-}\pm i0$ are possible
(zeros of $F(\cdot)$ on the imaginary axis) and the resonances are the zeros of $F(\cdot)$
on the lower half plane (fourth quadrant). For example in the case of the square well
potential there are at most finitely many poles on the rims $\Bbb{R}_{-}\pm i0$.
Also (iii) of Section 5 is satisfied (for details see e.g. [10]). The eigenspace
for the resonance $\zeta$ of
the transformed semigroup $t\rightarrow Re^{-itB_{+}}R^{\ast}$ is given by 
$\Bbb{C}e_{\zeta}$ where
\[
e_{\zeta}:=R\left\{\frac{1}{\cdot-\zeta}\right\},\quad  S(\overline{\zeta})=0.
\]

\vspace{3mm}

\subsection{Decay properties}
Recall that, without restriction of generality, $S(\cdot)$ can be considered
as the scattering matrix of the asymptotically complete scattering system
$\{H,M_{+}\}$ acting on ${\cal H}_{+}$ (see Section 2). That is, the quantum mechanical
evolution, restricted to the absolutely continuous subspace, is given by
the unitary evolution group 
\begin{equation}
\Bbb{R}\ni t\rightarrow e^{-itH}
\end{equation}
and the corresponding
"free" evolution by $t\rightarrow e^{-itM_{+}}$, both acting on ${\cal H}_{+}.$
On the other hand, as a counterpart, the set of resonances ${\cal R}$ causes the existence and
leads to the construction of the semigroup $\Bbb{R}_{+}\ni t\rightarrow
Re^{-itB_{+}}R^{\ast}$, acting on ${\cal H}_{+}$, too. It can be considered as the
{\em Decay Semigroup}, associated with the evolution (37), such that $\mbox{spec}\,B_{+}
={\cal R}$. However, this correspondence raises the problem to study the time dependence
of the semigroup compared to that of the free or unperturbed evolution in 
more detail, especially
in view of the transition probabilities of states.

For example, let
\[
e_{\zeta,k}(\lambda):=R\left\{\frac{k}{\cdot-\zeta}\right\},\quad S(\overline{\zeta})^{\ast}
k=0,
\]
be the transformed eigenvector of the eigenvector $\frac{k}{\cdot-\zeta}$ of $\zeta$
w.r.t. the semigroup $e^{-itB_{+}}$ on ${\cal T}\subset{\cal H}^{2}_{+}$. Then
\[
\Vert e_{\zeta,k}\Vert_{{\cal H}_{+}}=\left\Vert\frac{k}{\cdot-\zeta}\right\Vert_
{{\cal H}^{2}_{+}}.
\]
The transition probability w.r.t. the unperturbed evolution is
\[
\vert(e_{\zeta,k},e^{-itM_{+}}e_{\zeta,k})_{{\cal H}_{+}}\vert^{2}
=\left\vert\left(R\frac{k}{\cdot-\zeta},e^{-itM_{+}}R\frac{k}{\cdot-\zeta}\right)_
{{\cal H}_{+}}\right\vert^{2}=
\left\vert\left(\frac{k}{\cdot-\zeta},R^{\ast}e^{-itM_{+}}R\frac{k}{\cdot-\zeta}\right)_
{{\cal H}^{2}_{+}}\right\vert^{2},
\]
where $t\rightarrow R^{\ast}e^{-itM_{+}}R$ is the transform of the unperturbed evolution
from ${\cal H}_{+}$ onto the Hardy space ${\cal H}^{2}_{+}$.
On the contrary,
the transition probability w.r.t the decay semigroup is given by
\[
\vert(e_{\zeta,k},Re^{-itB_{+}}R^{\ast}e_{\zeta,k})_{{\cal H}_{+}}\vert^{2}=
\left\vert\left(\frac{k}{\cdot-\zeta},e^{-itB_{+}}\frac{k}{\cdot-\zeta}\right)_
{{\cal H}^{2}_{+}}\right\vert^{2}
\]
\[
=\left\vert\left(\frac{k}{\cdot-\zeta},e^{-itM}\frac{k}{\cdot-\zeta}\right)_
{\cal H}\right\vert^{2}
=\mbox{exp}(-2t\vert\mbox{Im}\,\zeta\vert)\Vert e_{\zeta,k}\Vert_{{\cal H}_{+}}^{2}.
\]
In the third term of this equation the unperturbed evolution $e^{-itM}$ appears, 
however w.r.t. the whole real line. That is, if the scattering matrix is a
univalent function of the energy parameter as a complex one (see Subsection 5.1),
i.e. if the unperturbed Hamiltonian $M_{+}$ can be extended to $M$ onto ${\cal H}$
then the transition probability w.r.t. the semigroup can be considered as usual, 
w.r.t. to the
extended evolution $e^{-itM}$. Essentially this is the case of the 
LP-theory (apart from the fact that the "classical" theory deals with only the case
that $S(\cdot)$ is holomorphic on the upper half plane which corresponds to the
orthogonality of the in/out-subspaces). 
In the multivalent case (see
Subsection 5.2) the problem is to compare the unitary evolution $e^{-itM_{+}}$ acting
on ${\cal H}_{+}$ with the evolution $Re^{-itB_{+}}R^{\ast}$, in particular the
transition probabilities of the eigenvectors $e_{\zeta,k}$ of the decay semigroup for
resonances $\zeta$, i.e. to estimate the difference
\[
\vert(e_{\zeta,k},e^{-itM_{+}}e_{\zeta,k})_{{\cal H}_{+}}\vert^{2}
-\mbox{exp}(-2t\vert\mbox{Im}\,\zeta\vert)\Vert e_{\zeta,k}\Vert_{{\cal H}_{+}}^{2},
\quad t>0,
\]
or for certain intervals of $t$.
The conceptual characterization of the set of all resonances as the spectrum of the
decay semigroup which is canonically associated to the scattering system
$\{H,M_{+}\}$ and to its scattering operator can be considered as a type of
{\em time-dependent characterization}. All the more such estimations are revealing,
however the presented characterization itself does not contribute to
this problem. 

\vspace{3mm}

REMARK. In the paper it is assumed that the multiplicity is finite. The proof of
similar results for the case $\dim{\cal K}=\infty$ requires additional considerations,
for example because in this case $\ker S(\overline{\zeta})^{\ast}=\{0\}$ does not
imply that this operator is bounded invertible. 

The properties of $B_{+}$ stated in
Theorems 2 and 3 are also true if $S(\cdot)$ has finitely many poles on
$\Bbb{C}_{+}$ as well as finitely many poles on $\Bbb{R}_{-}\pm i0$. In this case
one has to combine the arguments in the proofs of those theorems. The conjecture is that
the results are also true if there is a denumerable set of poles in $\Bbb{C}_{+}$.

\section{Trace class perturbations with analyticity conditions}

In this section a special class of trace class perturbations is presented such that
the assumptions of Theorem 3 are satisfied.

Let $V$ be a selfadjoint trace operator on ${\cal H}_{+}$ which is factorized by
\begin{equation}
V=BA^{\ast}=AB^{\ast}
\end{equation}
where $A$ and $B$ are Hilbert-Schmidt operators acting on an auxiliary 
Hilbert space ${\cal F}$. The image spaces $\mbox{ima}\,A$ and $\mbox{ima}\,B$ are
assumed to generate ${\cal H}_{+}$ w.r.t. $M_{+}$. $A$ and $B$ act by Hilbert-Schmidt
operator valued functions $A(\cdot),B(\cdot)$ from ${\cal F}$ into ${\cal K}$ by
\[
(Af)(\lambda):=A(\lambda)f,\quad (Bf)(\lambda):=B(\lambda)f,\quad f\in{\cal F}.
\]
Then
\[
\Vert A\Vert^{2}_{2}=\int_{0}^{\infty}\Vert A(\lambda)\Vert^{2}_{2}d\lambda,\quad
\Vert B\Vert^{2}_{2}=\int_{0}^{\infty}\Vert B(\lambda)\Vert^{2}_{2}d\lambda
\]
and
\[
\int_{0}^{\infty}\Vert A(\lambda)^{\ast}B(\lambda)\Vert_{1}\leq\Vert A\Vert_{2}\cdot
\Vert B\Vert_{2},
\]
where $\Vert\cdot\Vert_{1},\Vert\cdot\Vert_{2}$ denote the trace and Hilbert-Schmidt norm,
respectively. Then
\begin{equation}
T(z):=
A^{\ast}R_{0}(z)B=\int_{0}^{\infty}\frac{A(\lambda)^{\ast}B(\lambda)}{z-\lambda}d\lambda,
\quad z\in\Bbb{C}_{>0}:=\Bbb{C}\setminus[0,\infty),
\end{equation}
is a trace class valued holomorphic operator function, where $R_{0}(z)=(z\EINS-M_{+})^{-1}$.
Note that
\begin{equation}
(\EINS-A^{\ast}R_{0}(z)B)^{-1}=\EINS+A^{\ast}R(z)B,\quad z\in\Bbb{C}_{+}\cup\Bbb{C}_{-},
\end{equation}
where $R(z):=(z\EINS-H)^{-1}$, i.e. the left hand side is holomorphic
on $\Bbb{C}_{+}\cup\Bbb{C}_{-}$.

$\{H,M_{+}\}$ is an asymptotically complete scattering system. Its scattering matrix
can be calculated explicitly (see e.g. [19, p. 393]):
\begin{equation}
S(\lambda)=\EINS_{\cal K}-2\pi iB(\lambda)(\EINS+A^{\ast}R(\lambda+i0)B)A(\lambda)^{\ast},
\quad \lambda>0.
\end{equation}
In (41) $A$ and $B$ can be mutually replaced because of (38). Next an analyticity
condition is introduced. 
\begin{itemize}
\item[(i)]
The operator functions $A(\cdot),B(\cdot)$ are holomorphically continuable as Hilbert-Schmidt-
valued operator functions into $\Bbb{C}\setminus(-\infty,0]=:\Bbb{C}_{<0}$. On the
rims $\Bbb{R}_{-}\pm i0$ they are meromorphic with at most finitely many poles. Further
there is a region $G_{R,\epsilon}:=\{z\in\Bbb{C}_{<0}:\vert z\vert<\epsilon,
\vert z\vert>R\}$ where $R>\epsilon>0$ such that $\Vert A(\cdot)\Vert$ and $\Vert B(\cdot)
\Vert$ are bounded on this region, i.e. $\sup\{\Vert A(z)\Vert+\Vert B(z)\Vert\}<\infty$
for $z\in G_{R,\epsilon}$.
\end{itemize}
For later use we choose $R>\mbox{max}\{\vert\lambda\vert\}$ where $\lambda<0$ is a pole 
of $A(\cdot),B(\cdot)$ or a negative eigenvalue of $H$.
Then, according to (39), the operator function $z\rightarrow T(z)$
is holomorphically continuable across $\Bbb{R}_{+}$ into $\Bbb{C}_{-}$ from
$\Bbb{C}_{+}$ and into $\Bbb{C}_{+}$ from $\Bbb{C}_{-}$ as a trace class valued
operator function. 
For example, on $\Bbb{R}_{+}$ on has
\begin{equation}
T(\mu\pm i0)=\int_{0}^{\infty}\frac{A(\lambda)^{\ast}B(\lambda)}{\mu-\lambda}d\lambda\mp
i\pi A(\mu)^{\ast}B(\mu),
\end{equation}
where in this case the integral is Cauchy's mean value.
We put
\begin{equation}
L_{0}(z):=\EINS-A^{\ast}R_{0}(z)B,\quad z\in\Bbb{C}_{>0},
\end{equation}
The analytic continuation of $L_{0}(\cdot)$ across $\Bbb{R}_{+}$ from $\Bbb{C}_{\pm}$
into $\Bbb{C}_{\mp}$ is denoted by $L_{\pm}(\cdot).$ The corresponding "global" function
is denoted by $L(\cdot)$. It is holomorphic on its domain
${\cal D}:=\Bbb{C}_{>0}\cup(\Bbb{R}_{+}\cup\Bbb{C}_{-})\cup(\Bbb{R}_{+}\cup\Bbb{C}_{+})$.

(40) implies that $L(\cdot)^{-1}$ is meromorphic on ${\cal D}$ and a point $\zeta\in{\cal D}$
is a pole of $L(\cdot)^{-1}$ iff $\mbox{ker}\,L(\zeta)\supset\{0\}$ (see e.g. Gohberg/
Krein [21, p. 64]).
In particular, for $L_{0}(\cdot)^{-1}$ there are poles at most on $\Bbb{R}_{-}$ and
$\mu<0$ is a pole iff $\mbox{ker}\,L_{0}(\mu)\supset\{0\}$.

Furthermore, $S(\cdot)$ is analytically continuable into $\Bbb{C}_{<0}$ and one has
\begin{equation}
S(z)=\EINS-2\pi iB(z)L_{\iota}(z)^{-1}A(\overline{z})^{\ast},\quad z\in\Bbb{C}_{<0},
\end{equation}
where $\iota=0$ if $z\in\Bbb{C}_{+}$ and $\iota=+$ if $z\in\Bbb{R}_{+}\cup\Bbb{C}_{-}$.
It is holomorphic on $\Bbb{C}_{+}\cup\Bbb{R}_{+}$ and meromorphic on $\Bbb{C}_{-}$.
The poles can accumulate at most at $0$ and infinity. It is meromorphic on
$\Bbb{R}_{-}\pm i0$, too.

A further consequence of condition (i) is the absence of a singular continuous spectrum.

\subsection{Eigenvalues and resonances}

Obviously, the negative eigenvalues $\mu<0$ of $H$ can be characterized by $L_{0}$:
$\mu<0$ is an eigenvalue of $H$ iff $\mbox{ker}\,L_{0}(\mu)\supset\{0\}$. It is
well-known that condition (i) implies that this is also true for the positive
eigenvalues $\lambda>0$: it is an eigenvalue of $H$ iff $\mbox{ker}\,L_{+}(\lambda)
\supset\{0\}$, in this case the corresponding pole of $L_{+}(\cdot)^{-1}$ is also
simple. Interestingly enough, this characterization is also true for the
resonances, i.e. the poles of $S(\cdot)$ in $\Bbb{C}_{-}$ (see e.g. [19, p. 396])
\begin{itemize}
\item
The point $\zeta\in\Bbb{C}_{-}$ is a resonance iff it is a pole of
$L_{+}(\cdot)^{-1}$, i.e. iff $\mbox{ker}\,L_{+}(\zeta)\supset\{0\}$.
\end{itemize}
This expresses the close relation between eigenvalues of $H$ and resonances
of $\{H,M_{+}\}$, they can be described by a unique condition. In some sense it is a 
{\em spectral} characterization of the resonances in terms of $H$ (cf.
the corresponding remarks in Section 1). Note that a corresponding coincidence does 
not hold for the poles of $S(\cdot)$ on $\Bbb{R}_{-}+i0$. However one has: if there
are at most finitely many negative eigenvalues of $H$ then $S(\cdot)$ has at most
finitely many poles on $\Bbb{R}_{-}+i0$ (note (44) and condition (ii)).

\subsection{Boundedness properties of the scattering matrix}

For brevity we put $C(z):=A(\overline{z})^{\ast}B(z)$. A sufficient condition
such that the boundedness assumptions of $S(\cdot)$ in Theorem 3 are satisfied is given
by
\begin{itemize}
\item[(ii)]
The norm limit of $C(\cdot)$ for $z\rightarrow 0,\,z\in\Bbb{C}_{<0}$ exists uniformly and
vanishes. The integral
\[
\int_{0}^{\infty}\frac{C(\lambda)}{\lambda}d\lambda=-C_{0}
\]
exists as a compact operator, $\mbox{ker}\,(\EINS-C_{0})=\{0\}$ and the Cauchy mean value
$\int_{0}^{\infty}\frac{C(\lambda)}{\mu-\lambda}$ is norm convergent for $\mu\rightarrow 0$
with limit $C_{0}$.
\end{itemize}
Then the norm limit of $L(\zeta)$ for $\zeta\rightarrow 0$ exists
uniformly w.r.t. ${\cal D}$ and equals $(\EINS-C_{0})$. 
Its inverse $(\EINS-C_{0})^{-1}$ is bounded, i.e.
$0\in\mbox{res}\,(\EINS-C_{0})$. Then also $0\in\mbox{res}\,L(\zeta)$ is true  where
$\vert\zeta\vert$ is sufficiently small and $\sup\Vert L(\zeta)^{-1}\Vert<\infty$
if $\zeta$ varies in a small circle $\vert z\vert<\epsilon$. 
This gives $\sup\Vert S(z)\Vert<\infty$
for this circle. In particular, the number of negative eigenvalues of $H$ is finite.

Further let $Q:=\{z:\vert\mbox{Re}\,R\vert<a,0\leq\mbox{Im}\,z<b\}$ be a rectangle such
that $G:=\Bbb{C}_{+}\setminus Q\subset G_{R,\epsilon}$. Then
\begin{equation}
\sup_{z\in G}\Vert S(z)\Vert<\infty.
\end{equation}
If $\mbox{Re}\,z\leq-a$ or $\mbox{Im}\,z\geq b$ then (45) is obvious. The function
\[
F(z):=B(z)(\EINS+A^{\ast}R(z)B)A(\overline{z})^{\ast}=\frac{1}{2\pi i}(\EINS-S(z))
\]
is holomorphic in the rectangle $R\leq\mbox{Re}\,z\leq X,0\leq\mbox{Im}\,z\leq b$
including its boundary. Then from the maximum principle for holomorphic functions
and the fact that this maximum is independent of $X$ (45) follows.

That is, for trace class perturbations satisfying the conditions (i) and (ii) the
assumptions of Theorem 3 are satisfied. Note that for this class the assumption
$\mbox{dim}\,{\cal K}<\infty$ is dispensable because in this case
$S(z)-\EINS$ is trace class such that $\mbox{ker}\,S(\overline{\zeta})^{\ast}=\{0\}$
implies that $(S(\overline{\zeta})^{\ast})^{-1}=S(\zeta)$ is bounded. 

Also the class of generalized Friedrichs models (see [11]) contains examples which satisfy
the assumptions of Theorem 3.

\subsection{An example}

Let $\dim\,{\cal K}=1$, i.e. the Hilbert space is $L^{2}(\Bbb{R}_{+},\Bbb{C},d\lambda)$.
Put $V=aP$, where $P:=(e,\cdot)e,\,a\neq 0$. Choose
\[
e(\lambda):=\sqrt{\frac{2}{\pi}}\cdot\frac{\lambda^{1/4}}{\lambda+1},\quad \Vert e\Vert=1.
\]
The scattering matrix is given by
\[
S(\lambda)=1-2\pi ia\cdot e(\lambda)(1-a(e,R_{0}(\lambda+i0)e))^{-1}\overline{e(\lambda)}
\]
The calculation of $(e,R_{0}(z)e)$ gives
\[
(e,R_{0}(z)e)=-\frac{1}{(1-iz^{1/2})^{2}},\quad \mbox{Im}\,z^{1/2}>0.
\]
Then the scattering matrix reads
\begin{equation}
S(\lambda)=1-4ia\frac{\lambda^{1/2}}{(1+i\lambda^{1/2})^{2}(a+(1-i\lambda^{1/2})^{2})}
\end{equation}
That is, the Riemann surface for $S(\cdot)$ is that of $z^{1/2}$. We put
$z^{1/2}=:k$. Then $k$ varies over the whole complex plane. The eigenvalue equation reads
\[
(1-ik)^{2}+a=0.
\]

(i) $a>0$. There are two solutions, both in the second sheet:
\[
\zeta=k^{2}=a-1\pm 2i\sqrt{a},
\]
i.e. one obtains a resonance and the complex conjugated anti-resonance.

(ii) $a<0$. There are two solutions:
\[
k_{1}:=-i(1+\sqrt{-a}),\quad k_{2}:=-i(1-\sqrt{-a}).
\]
If $a<-1$ then $\mbox{Im}\,k_{1}<0$, i.e. $k_{1}^{2}\in\Bbb{R}_{-}$ is from the second sheet
and $\mbox{Im}\,k_{2}>0$, i.e. $k_{2}^{2}\in\Bbb{R}_{-}$ is from the first sheet, it
is an eigenvalue.

If $-1<a<0$ then both solutions are in $\Bbb{R}_{-}$ and from the second sheet. 

That is, for $a>0$ there is no negative eigenvalue, $S(\cdot)$ has no pole on $\Bbb{R}_{-}$,
there is one resonance, the scattering matrix is given by (46),
it is bounded near $k=0$ and bounded for sufficiently large $\vert k\vert$.

\section{Acknowledgement}
It is a pleasure to thank K.B. Sinha for a stimulating discussion at the
XXV WGMP in Bialowieza 2006 and L.S. Schulman and A. Bohm for inviting me to participate
at the Advanced Study Group 2008 on "Time: Quantum and Statistical Mechanics Aspects", 
held at the Max-Planck-Institute for the Physics of Complex Systems in Dresden,
where I gave a talk on this topic, and for sponsoring my participation.

\section{References}

\begin{enumerate}
\item
Bohm, A.: Quantum Mechanics,\\
Springer Verlag Berlin 1979
\item 
Reed, M. and Simon, B.: Methods of Modern Mathematical Physics IV:\\ Analysis of Operators\\
Academic Press New York San Francisco London 1978
\item
Br\"andas, E. and Elander, N. (eds.): Resonances,\\
Lecture Notes in Physics 325, Springer Verlag Berlin 1989
\item
Aigular, J. and Combes, J. M.: A class of analytic perturbations for one-body Schr\"odinger 
Hamiltonians\\
Commun. Math. Phys. 22, 269 - 279 (1971)
\item
Simon, B.: Resonances in N-body quantum systems with dilatation analytic potentials
and the foundations of time-dependent perturbation theory\\
Ann. Math. 97, 247 - 274 (1973)
\item
Hislop, P. D. and Sigal, I. M.: Introduction to Spectral Theory: With Applications to
Schr\"odinger Operators\\
Springer Verlag 1996
\item
Howland, J.W.: On the Weinstein-Aroszajn formula,\\
Arch. rat. Mech. Anal. 39, 323 - 339 (1970)
\item
Baumg\"artel, H.: Resonances of Perturbed Selfadjoint Operators and their Eigenfunctionals,\\
Math. Nachr. 75, 133 - 151 (1976)
\item
Bohm, A. and Gadella, M.: Dirac kets, Gamov vectors and Gelfand triplets,\\
Lecture Note in Physics 348, Springer Verlag 1989
\item
Baumg\"artel, H, Kaldass, H. and Komy, S.: Spectral theory for resonances of
real-valued central-symmetric potentials with compact support\\
Journal Math. Phys. 50, Nr. 2 (2009)
\item
Baumg\"artel, H.: Spectral and Scattering Theory of Friedrichs Models on the Positive
Half Line with Hilbert-Schmidt Perturbations,\\
Ann. Henri Poincare' 10, 123 - 143 (2009)
\item
Lax, P.D., Phillips, R.S.: Scattering Theory,\\
Academic Press, New York 1967
\item
Strauss, Y., Horwitz, L.P. and Eisenberg, E.: Representation of quantum mechanical
resonances in the Lax-Phillips Hilbert space,\\
Journal Math. Phys. 41, 8050 (2000)
\item
Strauss, Y., Horwitz, L.P. and Eisenberg, E.: The Lax-Phillips Semigroup of the Unstable
Quantum System,\\
Lecture Notes in Physics 504, 323 - 332 (1998)
\item
van Winter, C.: J. Math. Anal. 47, 633 (1974)
\item
Kato,T.: Perturbation Theory for Linear Operators,\\
Springer Verlag Berlin 1976
\item
Halmos, P.R.: Two Subspaces,\\
Trans. Amer. Math. Soc. 144, 381 - 389 (1969)
\item
Wollenberg,M.: On the inverse problem in the abstract theory of scattering,\\
ZIMM-Preprint Akad. Wiss. DDR, Berlin 1977
\item
Baumg\"artel, H., Wollenberg, M.: Mathematical Scattering Theory,\\
Birkh\"auser Basel Boston Stuttgart 1983
\item
Baumg\"artel, H.: On Lax-Phillips semigroups,\\
J. Operator Theory 58, 23 - 38 (2007)
\item
Gohberg, I.C. and Krein, M.G.: The basic propositions on defect numbers,
root numbers and indices of linear operators,\\
Uspechi Mat. Nauk 12, 2 (74), 43 - 118 (1957) (Russian)

\end{enumerate}
\end{document}